\documentclass[preprint2]{aastex}
\def\stacksymbols #1#2#3#4{\def\theguybelow{#2}
	\def\verticalposition{\lower#3pt}
	\def\spacingwithinsymbol{\baselineskip0pt\lineskip#4pt}
	\mathrel{\mathpalette\intermediary#1}}
\def\intermediary #1#2{\verticalposition\vbox{\spacingwithinsymbol
	\everycr={}\tabskip0pt
	\halign{$\mathsurround0pt#1\hfil##\hfil$\crcr#2\crcr
		\theguybelow\crcr}}}
\def\lta{\stacksymbols{<}{\sim}{2.5}{.2}}
\def\gta{\stacksymbols{>}{\sim}{3}{.5}}

\begin{document}

\title{WHERE DO COOLING FLOWS COOL?}



%



\author{Fabrizio Brighenti$^{1,2}$ and William G. Mathews$^1$}

\affil{$^1$University of California Observatories/Lick Observatory,
Board of Studies in Astronomy and Astrophysics,
University of California, Santa Cruz, CA 95064\\
mathews@lick.ucsc.edu}

\affil{$^2$Dipartimento di Astronomia,
Universit\`a di Bologna,
via Ranzani 1,
Bologna 40127, Italy\\
brighenti@bo.astro.it}






\vskip .2in

\begin{abstract}

Typically $\sim 5$ percent of the total baryonic mass in luminous
elliptical galaxies is in the form of cooled interstellar gas.
Although the mass contributed by cooled gas is small relative to the
mass of the old stellar system in these galaxies, it is almost
certainly concentrated within the optical effective radius where it
can influence the local dynamical mass.  However, the mass of cooled
gas cannot be confined to very small galactic radii ($r \lta 0.01r_e$)
since its mass would greatly exceed that of known central mass
concentrations in giant ellipticals, normally attributed to massive
black holes.  We explore the 
proposition that a population of very low mass, optically
dark stars is created from the cooled gas.  For a wide variety of
assumed radial distributions for the interstellar cooling, we find
that the mass of cooled gas contributes significantly ($\sim$30
percent) to stellar dynamical mass to light ratios which, as 
a result, are expected to vary with galactic radius.  
However, if the stars formed from cooled interstellar gas are 
optically luminous, their pertubation on the the mass to light 
ratio of the old stellar population may be reduced.
Cooling mass dropout also
perturbs the local apparent X-ray surface brightness distribution,
often in a positive sense for centrally concentrated cooling.  In
general the computed X-ray surface brightness exceeds observed values
within $r_e$, suggesting the presence of additional support by
magnetic stresses or non-thermal pressure.  The mass of cooled gas
inside $r_e$ is sensitive to rate that old stars lose mass ${\dot
M_*}$, but this rate is nearly independent of the initial mass
function of the old stellar population.

\end{abstract}

\keywords{galaxies: elliptical and lenticular -- 
galaxies: evolution --
galaxies: cooling flows --
galaxies: interstellar medium --
x-rays: galaxies}

\section{INTRODUCTION}

Perhaps the most perplexing and long-standing problem 
associated with galactic and cluster cooling flows is the 
uncertain physical nature and spatial distribution 
of the gas that cools. 
The apparent absence of large 
masses of cooled gas in elliptical 
galaxies has led some 
to argue that little or no cooling actually occurs  
and to postulate some source of heating that offsets 
the radiative losses in X-ray emission. 
But the energy required to balance 
radiative losses is prohibitively large and 
appropriate heating sources may not be universally 
available.
If the expected radiative cooling actually occurs, 
two questions arise:
(1) What is the nature of the objects 
that condense from the hot gas? and 
(2) Where is most of the cooled mass located 
in the galaxy?
Regarding the first question, 
a variety of physical arguments support the 
hypothesis, or even the inevitability, 
of low mass star formation 
(Fabian, Nulsen \& Canizares 1982; 
Thomas 1986;
Cowie \& Binney 1988;
Vedder, Trester, \& Canizares 1988;
Sarazin \& Ashe 1989;
Ferland, Fabian \& Johnstone 1994; 
Mathews \& Brighenti 1999).
Here we shall address the second question in the 
context of cooling flows in elliptical galaxies where 
the known stellar mass and light profiles 
strongly constrain the spatial distribution of cooled gas.

We adopt the generally accepted hypothesis that only 
stars of very low mass (e.g. $\lta 0.1$ $M_{\odot}$) 
form in cooling flows (e.g. Ferland, Fabian \& Johnstone 1994), 
so that the mass to light ratio of the young stellar population 
formed from the cooling gas is essentially infinite.
In view of the difficulties we encounter with this 
hypothesis, described below, 
it seems more likely that the stellar population 
formed from cooled gas extends to somewhat more 
massive stars that are optically luminous.

Our gas dynamical models for the evolution of hot interstellar 
gas in giant ellipticals indicate that the origin 
of the gas varies with galactic radius.
Most of the gas in the inner, optically luminous 
regions originates from the ejected envelopes of 
evolving stars; gas in the outer halo is supplied by 
cosmological secondary infall or tidal acquisitions from 
neighboring galaxies (Mathews \& Brighenti 1998b). 
Circumgalactic gas around massive ellipticals is enriched  
by Type II supernovae that accompanied early star formation. 
The variability of circumgalactic gas among luminous 
ellipticals is responsible for some of the 
enormous dispersion in X-ray luminosity $L_x$ 
among ellipticals of similar 
optical luminosity $L_B$ (Mathews \& Brighenti 1998a).

Since the hot interstellar gas in a bright elliptical 
emits observable X-rays,
it is clearly losing energy.
However, as the gas loses energy 
it is compressed toward the galactic center by gravitational 
forces and $Pdv$ work maintains the high temperatures observed,
$T \sim 10^7$ K, producing a galactic cooling flow. 
The positive interstellar temperature gradients typically observed
within
a few effective radii are often cited as evidence of radiative
cooling in a cooling flow,
but this cooling is due instead to the mixing
of somewhat cooler, locally virialized gas ejected from stars
with hotter gas arriving from larger galactic
radii (Mathews \& Brighenti 1998b; Brighenti \& Mathews 1998, 
1999a).
If large entropy fluctuations are present in the hot 
gas, catastrophic 
cooling can occur at any radius in the flow.
Regions of low entropy (low temperature, high density) 
radiate more and cool sooner. 
The amplitude distribution of entropy fluctuations 
in the interstellar gas 
determines the radius 
where cooling mass dropout occurs in the cooling flow. 
For example, if the entropy in some region in the flow 
is only slightly less than in the ambient flow,
the differential radiative cooling will be 
small and the region will cool out of the flow 
at small galactic radii; conversely, localized regions with 
entropy much lower than the ambient flow 
cool rapidly and deposit their mass at large radii.
Some possible sources of interstellar entropy variations are 
stellar winds, 
explosions of Type Ia supernovae, non-uniform SNII heating 
at early times, and 
mergers with small, gas-rich galaxies. 

The total rate that mass cools and drops out of the flow 
is closely related to the X-ray luminosity $L_x$.
The X-ray luminosity can be approximately expressed as the 
product of the total cooling rate ${\dot M}$ and the 
enthalpy per gram in the hot gas, or 
$${\dot M} = \left({2 \mu m_p \over 5 k T}\right) L_{x,bol}
\approx 2.5 M_{\odot} {\rm yr}^{-1}.$$
Here we have used data from the giant Virgo elliptical 
NGC 4472: $T \approx 1.3 \times 10^7$ K; 
$L_x(0.5 - 4.5 ~{\rm keV}) = 4.5 \times 10^{41}$;
$L_{x,bol} \approx 1.6 L_x(0.5 - 4.5 ~{\rm keV})$.
If $L_x$ and $T$ are reasonably constant over the Hubble time, 
a mass $M_{cg} \approx 3 \times 10^{10}$ 
$M_{\odot}$ of cold gas is expected to condense from the 
hot ISM somewhere within NGC 4472.
Although this mass is very large, it is only about 4 percent 
of the total stellar mass in NGC 4472 today.
The mass that cools can therefore be ignored 
if it is widely distributed throughout the galactic volume.
However, the central concentration of H$\alpha$ emission 
in ellipticals (e.g. Macchetto et al. 1996) 
suggests that the cooling 
is concentrated toward the galactic 
center where the interstellar density is highest 
and the bulk of the X-ray energy is emitted. 

The motivation of this paper is to explore 
a variety of options for the mass dropout profile of cooled 
gas in bright ellipticals appropriately constrained by 
the known radial distributions of total stellar and 
non-baryonic mass. 
The radial mass dropout profile of cooled gas 
cannot be determined from first principles 
because the distribution and amplitude of 
the entropy and magnetic fluctuations in the hot gas
are unknown and difficult to evaluate 
from simple physical arguments.
Nevertheless, the total mass of cooled gas inside an 
effective (half-light) radius $r_e$ must be consistent 
with the mass to light ratio determined from stellar velocities 
and with the total mass 
inferred from X-ray observations within $r_e$. 
Assuming that the stellar mass to light ratio is 
uniform with radius, the stellar mass $M_*(r)$ and the 
X-ray mass $M_x(r)$ appear to be in nearly perfect agreement 
for two bright Virgo ellipticals 
in the range $0.1r_e \lta r \lta r_e$ 
(Brighenti \& Mathews 1997a). 
Because of 
constraints on the mass distribution of cooled gas 
provided by X-ray and stellar dynamical observations,
galactic cooling flows provide a critical venue for testing 
the physics of mass deposition in cooling flows.

The mass of cold ($T \lta 10^4$ K) gas $M_{cg}$ 
actually observed 
in ellipticals is many orders of magnitude less than 
the total cooled mass estimated above. 
For example, neither HI nor H$_2$ gas has been observed in NGC 4472, 
only upper limits, $M_{cg} \lta 10^7$ $M_{\odot}$
(Bregman, Roberts \& Giovanelli 1988;
Braine, Henkel \& Wiklind 1988). 
If stars form they must either be indistinguishable from 
the old stellar population or non-luminous.
We explore here the possibility that most of the 
cooled gas forms dense baryonic clouds or stars 
that are dark at optical and radio frequencies.
The very low mass stars advanced by 
Ferland, Fabian \& Johnstone (1994) 
satisfy this invisibility criterion, 
while the star formation models of  
Mathews \& Brighenti (1999) indicate that 
(luminous) stars of mass $\sim 1 - 2$ $M_{\odot}$ can form 
in galactic cooling flows.

X-ray studies indicate significant masses of 
cold, absorbing gas in cluster and galactic cooling flows 
(e.g. White et al. 1991; Allen et al. 1993;
Fabian et al. 1994; Allen \& Fabian 1997; Buote 1999), 
but these results are inconsistent with 
the absence of radio frequency emission 
from the cold gas (Braine \& Dupraz 1994; 
Donahue \& Vogt 1997) and should 
be regarded as controversial until this inconsistency is resolved. 
Even taken at face value, the total mass of cooled gas implied 
by the X-ray absorption in cluster cooling flows 
is typically only a small fraction of the 
total mass that should have cooled in a Hubble time
(Allen \& Fabian 1997;
Wise \& Sarazin 1999), implying that most of the 
cooled gas may have formed stars.   
If cooled gas forms into small stars,
these stars will have apogalatica near their point of origin 
where they will spend most of their orbital time. 
We shall assume that the gas mass that cools and drops out of the 
flow contributes optically dark (stellar)  
mass at the radius where the cooling occurred. 

In the following we describe 
a series of gas dynamical calculations for the evolution of  
X-ray emitting interstellar gas over the Hubble time 
and investigate a variety 
of assumptions about the radial distribution of 
optically dark cooled gas.
To be specific,
we compare our models with 
the well-observed massive elliptical NGC 4472.
We find that the mass of cooled gas contributes 
significantly to dynamical mass to light determinations 
within $r_e$ based on stellar velocities. 

\section{KNOWN STELLAR AND DARK MASS DISTRIBUTION IN NGC 4472}

The E2 elliptical NGC 4472 is a luminous, slowly rotating 
[$(v/\sigma)_* = 0.43$] galaxy in the Virgo cluster. 
With an adopted distance $d = 17$ Mpc, its optical luminosity 
is $L_B = 7.89 \times 10^{10}$ $L_{B\odot}$ and its 
half light or effective radius 
$r_e = 1.733'$ is $8.57$ kpc (Faber et al. 1989).
The total stellar mass 
$M_{*t} = 7.26 \times 10^{11}$ $M_{\odot}$ is found 
from the 
mass to light ratio $M/L_B = 9.2$ determined by 
van der Marel (1991) with a two-integral 
stellar distribution function.
This mass to light ratio is appropriate to the 
galactic region within about 0.4$r_e$ where 
stellar velocities are well determined, 
although the mass determined from X-ray observations 
suggests that $M/L_B$ remains constant to at least $r_e$
(Brighenti \& Mathews 1997a).
If $M/L_B$ is spatially constant, the stellar 
mass also has a de Vaucouleurs profile.
Within a central core or break radius 
$r_b = 2.41''$ ($200$ pc) the stellar density 
profile flattens (Faber et al. 1997), but 
we shall not consider this small feature here.
NGC 4472 contains a central black 
hole of mass $M_{bh} = 2.9 \times 10^9$ $M_{\odot}$
(Magorrian et al. 1998).

The total mass distribution in luminous ellipticals can 
most easily be determined 
from the radial variation of density 
and temperature in the  
hot interstellar gas, assuming hydrostatic equilibrium.
Figure 1 illustrates the interstellar density and temperature 
profiles in NGC 4472.
The filled circles in Figure 1 
are {\it Einstein} HRI observations 
(Trinchieri, Fabbiano \& Canizares 1986) and 
open circles are ROSAT HRI and PSPC data 
from Irwin \& Sarazin (1996) 
(also see Forman et al. 1993).
The $T(r)$ and $n(r)$ profiles have been fit with 
analytic curves as described by Brighenti \& Mathews 
(1997a).

Hydrostatic equilibrium in the hot interstellar gas 
is an excellent approximation 
since the cooling flow velocity is very subsonic. 
The total mass interior to radius $r$ determined from X-ray 
observations is 
\begin{equation}
M_{x}(r) = - {k T(r) r \over G \mu m_p }
\left( {d \log \rho \over d \log r} + {d \log T \over d \log r}
+ {P_m \over P} ~ {d \log P_m \over d \log r} \right)
\end{equation}
where $m_p$ is the proton mass and $\mu = 0.61$ is the 
mean molecular weight.
The last term, representing the possibility of
magnetic pressure $P_m = B^2/8 \pi$, is negative
if $d P_m / dr < 0$ as seems likely.
If the magnetic term is important but not included, 
the total mass will be underestimated. 
Assuming $P_m = 0$,
the total mass $M_x(r)$ in NGC 4472 is shown 
with a solid line in Figure 1b.

In the outer halo, $r \gta r_e$,
the total mass is dominated by the dark halo. 
The dark halo mass distribution in NGC 4472 
can be approximated with an NFW halo profile 
(Navarro, Frenk \& White 1996) 
of virial mass $M_h = 4 \times 10^{13}$ $M_{\odot}$ 
although the 
observed halo is somewhat less centrally peaked 
than NFW (Brighenti \& Mathews 1999a).
Within $r_e$ the contribution of the dark 
NFW halo mass 
in the model is small;
for example at $r < r_e/3$ the total mass to light ratio 
is $M/L_B = 10.18$, only 10 percent greater than the dynamic 
value 9.2 determined in $r \lta 0.4r_e$. 

It is remarkable that  
the total mass found from the X-ray data $M_x(r)$ 
is nearly identical to the 
expected dynamical mass $M_*(r)$ 
(based on stellar velocities and $M/L$)
in the range $0.1r_e \lta r \lta r_e$ (Figure 1).
An almost identical agreement in this radius range is 
indicated by X-ray observations of another bright 
Virgo elliptical, NGC 4649 (Brighenti \& Mathews 1997a). 
In this important region the 
hot gas is confined by the {\it stellar} potential.
The excellent agreement of the stellar and X-ray masses  
supports the  
consistency of two radically different mass determinations: 
from stellar velocities and from the radial equilibrium of 
hot interstellar gas.
The apparent agreement of the X-ray and stellar 
masses in this range of galactic radii 
also indicates that the hydrostatic 
support of the hot gas is not strongly influenced 
by local magnetic fields and rotation.

However, it is not obvious why $M/L_B$ would be constant 
with galactic radius, particularly when the cooling 
dropout mass is considered, and why the 
agreement between stellar dynamic and 
X-ray masses no longer obtains in the 
central regions $r \lta 0.1 r_e$. 
In this central region the total mass 
indicated by the X-ray observations in Figure 1 
is considerably 
less than the expected mass based on an assumed 
de Vaucouleurs profile and constant mass to light ratio. 
This type of deviation could be due to 
magnetic or other non-thermal pressure in this region,
to rotation, or 
to local cooling dropout in the hot interstellar gas. 
The lower mean temperature in cooling regions lowers
the total apparent gas temperature and results in 
an underestimate of the total internal mass 
(equation 1). 

We assume that currently available 
{\it Einstein} and ROSAT observations are accurate 
in the central region of NGC 4472, $r \lta 0.1 r_e$. 
These observations have been reduced assuming  
no (non-Galactic) photoelectric absorption by low temperature 
gas in the central regions.
If X-ray absorption is  
present, the true hot gas density would be 
more centrally peaked than 
shown in Figure 1 and 
the total mass indicated by the X-ray observations would 
increase.
Buote (1999) finds that two-temperature models fit the 
X-ray spectrum for NGC 4472 quite well.
The two temperatures do not necessarily need to be 
spatially mixed; they could also approximately represent 
the range of the radial 
temperature variation observed in NGC 4472. 
In Buote's two temperature model, only the cooler component  
($T \sim 0.7$ keV located in $r \lta r_e$) 
requires an absorption column $N_H = 2.9 \times 10^{21}$ 
cm$^{-2}$ in excess of the Galactic value.
However, the influence of cold gas having this column density 
on the hot gas density plotted in Figure 1 is small.
For a worst case example, suppose that absorbing material 
with column density $N_H = 2.9 \times 10^{21}$
cm$^{-2}$ 
is in a disk oriented perpendicular to the line of sight 
and that this disk 
absorbs {\it all} X-rays from the back side of the galaxy. 
The radius of this disk would be $< 370$ pc if it contained 
the maximum mass $M_{cg} \sim 10^7$ $M_{\odot}$ 
allowed by CO and HI observations 
(Bregman, Roberts \& Giovanelli 1988;
Braine, Henkel \& Wiklind 1997) 
or 20 kpc if it contained 
all of the gas that has cooled, $M_{cg} = 3 \times 10^{10}$
$M_{\odot}$.
The X-ray surface brightness 
within the opaque disk would be reduced by 2, 
but the corresponding gas density would be lowered by only 
$2^{1/2} = 10^{0.15}$ since the volume emissivity 
$\propto n^2$. 
Such a small correction in the density (gradient)  
could not account for the large mass 
discrepancy between the X-ray mass and the  
stellar dynamical mass shown in Figure 1 within 0.1$r_e$.

The densities and temperatures in Figure 1 
were determined from X-ray data 
assuming the abundance of the hot 
gas is uniformly solar. 
Since the gas is likely to be more metal rich at smaller 
galactic radii (Matsushita 1997; Brighenti \& Mathews 1999b),
an allowance for this gradient would tend to lower 
the derived density gradient and the internal mass,  
{\it increasing} the discrepancy in $r \lta 0.1~r_e$ 
in Figure 1 by a small amount.
In the following discussion we shall ignore the 
relatively small possible influence of absorption or metallicity 
gradients on the results shown in Figure 1.

\section{HYDRODYNAMICAL MODELS}

The hydrodynamical models we use in this paper are similar 
to those in our recent papers (e.g. Brighenti \& Mathews 1999a) 
so we provide only a brief review here. 
Hot interstellar gas in ellipticals 
has a dual origin: (i) mass loss from an evolving old stellar 
population and (ii) secondary infall into the overdensity 
perturbation that formed the galaxy group within which 
the elliptical formed by early merging events.
For a given set of cosmological parameters, dark 
and baryonic matter flow toward an overdensity region.
The dark matter forms an NFW halo, growing in size with 
time.
Spherical geometry is assumed.
Within the accretion shock at time $t_* = 2$ Gyr, 
when enough baryons have accumulated,
we form the current de Vaucouleurs stellar profile and 
release the energy of all Type II supernovae according 
to a Salpeter IMF (slope: $x = 1.35$, mass limits: 
$m_{\ell} = 0.08$ and $m_u = 100$ $M_{\odot}$).
All stars greater than 8 $M_{\odot}$ produce 
Type II supernovae each 
of energy $E_{sn} = 10^{51}$ ergs.
We assume that 
a fraction $\epsilon_{sn} = 0.8$ of this energy is 
delivered to the internal energy of the gas.
We have shown (Brighenti \& Mathews 1999b) that 
such a galaxy formation scheme can work in a variety of 
cosmologies: flat or low density, 
with or without a lambda term. 
The evolution of gas within the optical
effective radius $r_e$, of most interest 
here, is insensitive to these cosmological parameters. 
For simplicity therefore we 
assume a simple flat universe with $\Omega = 1$, 
$H_o = 50$ km s$^{-1}$ 
Mpc$^{-1}$ and $\Omega_b = 0.05$.
We characterize the dark halo with an 
NFW profile (Navarro, Frenk \& White 1996) having a 
virial mass $M_h = 4 \times 10^{13}$ $M_{\odot}$ 
at the current time $t_n = 13$ Gyrs.
The models we discuss here are identical to the 
standard model of (Brighenti \& Mathews 1999a) except 
we now use a finer central spatial zoning (65 pc for the 
innermost zone), a SNII efficiency $\epsilon_{sn} = 0.8$, 
and a mass ``dropout'' function $q(r)$ with more 
adjustable parameters (see below).

Our objective is to seek 
solutions of the gasdynamical equations 
including mass dropout that jointly satisfy several 
observational constraints at time $t_n$:
(i) the observed hot gas density, temperature 
and X-ray surface brightness profiles,
(ii) the known dynamical mass in the galactic center 
usually attributed to a massive black hole,
(iii) the apparent dynamical mass to light ratio 
$M/L_B = 9.2$ determined in $r \lta 0.4r_e$,
and
(iv) an apparent internal mass $M_x(r)$ in $(0.1 - 1)r_e$ 
based on Equation (1) that agrees with the 
constant-$(M/L_B)$ de Vaucouleurs profile as shown in 
Figure 1. 

The baryonic component in our models has a complex evolution.
Much of the initial 
baryonic mass is consumed in creating the stellar system. 
When the Type II supernova energy is released, 
a significant mass of gas is expelled as a galactic wind.
After these early events, the 
interstellar gas is re-established and sustained 
by stellar mass loss 
and by inflow of circumgalactic gas (secondary infall), most of 
which was previously enriched and expelled by SNII. 
We assume that the stars form during a short epoch that 
can be described by a single burst Salpeter IMF 
as discussed above.
The stellar mass loss rate for this IMF 
varies as 
$\dot{M_{*t}} = \alpha_*(t) M_{*t}$ where
$\alpha_*(t) = 4.7 \times 10^{-20} 
[t/(t_n - t_{*s})]^{-1.3}$ sec$^{-1}$.
Although galactic stars form at $t_{*s} = 1$ Gyr, 
their mass loss contribution to the ambient 
interstellar gas is assumed to begin at a later time,
$t_* = 2$ Gyrs.
Since galactic stars have been enriched 
by supernova ejecta, the single burst model cannot 
be strictly correct, but our approximation 
$\alpha_*(t < t_* ) = 0$ is 
consistent with several early 
starbursts closely spaced in time and allows for metal
enrichment of old galactic stars that are not in
the first single-burst population.
If the de Vaucouleurs profile is a result of 
largely dissipationless merging, 
some or most of the 
star formation must have occurred at a time $t_{*s}$ 
before the important merging events at $t_* = 2$ Gyrs.
By taking $t_{*s} < t_*$ we 
reduce by $\sim 10^{10}$ $M_{\odot}$ 
the total amount of gas ejected by stars 
within the galactic potential. 
We recognize the inconsistencies in these approximations 
of complex stellar formation and dynamical processes
that are poorly understood. 
However, once the galaxy is formed, we follow the 
interstellar gas dynamical evolution in full detail,
conserving mass and energy.

Continued heating by Type Ia supernova is assumed to vary
inversely with time, SNu$(t) =~$ SNu$(t_n)$$(t_n/t)$, 
where the current rate, SNu$(t_n) = 0.03$ SNIa 
per 100 yrs per $10^{10}$ $L_{B\odot}$, is near 
the lower limit of observed values,
SNu$(t_n) = 0.06 \pm 0.03(H/50)^2$ (Cappellaro et al. 1997), 
as required to maintain the low interstellar 
iron abundance. 

For the models discussed here 
the equation of continuity includes a ``mass dropout'' term:
$${ \partial \rho \over \partial t}
+ {1 \over r^2} { \partial \over \partial r}
\left( r^2 \rho u \right) = \alpha \rho_*
-q(r) {\rho \over t_{do}},$$
where $t_{do} = 5 m_p k T / 2 \mu \rho \Lambda$
is the time for gas to cool locally by radiative 
losses at constant pressure (see, e. g. Sarazin \& Ashe 1989).
The cooling is assumed to be instantaneous 
without advection in the cooling flow, 
i.e. $t_{do} \ll t_{flow} = r/v$, 
although in practice 
this inequality may not always be satisfied. 
While this type of cooling dropout has been widely used in 
past models, there is no adequate physical 
model for mass dropout. 
Clearly, the gas must cool somewhere -- the emission 
of X-rays indicates a large net energy (and mass) loss 
from the interstellar medium. 

When small regions of low entropy cool, the pressure 
remains constant since the sound crossing time 
is much less than the flow time. 
Following Fabian, Nulsen \& Canizares (1982) and  
Ferland, Fabian, \& Johnstone (1994), we assume that 
cooled gas converts to a second population of 
optically dark, low mass stars. 
Regarding H$\alpha$ emission as a tracer for 
the cooling gas,
Mathews \& Brighenti (1999) have shown that 
the cooling occurs at a multitude ($\sim 10^6$) of
cooling sites distributed throughout the inner galaxy 
and that
only stars with mass $\lta 1 - 2 M_{\odot}$ can form at each 
cooling site. 
This is supported by the observed absence of 
young massive stars and SNII in elliptical galaxies.
Nevertheless, in the following discussion 
we assume that the maximum stellar mass 
in the dropout population is sufficiently 
low that the optical light from these stars is 
unobservable.
We expect cooling and 
low mass star formation to be concentrated 
within at least 2 kpc,
which is the observed extent of H$\alpha$ emission 
in NGC 4472 (Macchetto et al. 1996).  
While the region containing optical emission lines 
provides a natural  
guideline for selecting the dropout profile,
we consider a wider range of constant or variable 
dropout coefficients $q(r)$ parameterized by 
\begin{equation}
q(r) = q_o \exp(-r/r_{do})^m
\end{equation}
which concentrates the cooling within radius $r_{do}$. 
Note that even when $q$ is constant, the mass dropout 
term is spatially concentrated, $\rho / t_{do} \propto 
\rho^2$.

\section{MODELS WITH ZERO OR CONSTANT q}

In a series of recent papers in this Journal 
we have presented evolutionary cooling flow models 
for NGC 4472 that agree quite well with 
the observed distributions 
of interstellar gas density, temperature and metallicity 
(Brighenti \& Mathews 1999a; 1999b), 
particularly at intermediate and large radii.
Since the total mass of cooled gas in these calculations 
was only a few percent of the stellar mass, 
the gravitational contribution of cooled gas was ignored,
even in models including mass dropout.
For a better understanding of the inner galaxy,
we now include the gravity of all gas, 
both hot and cold, except when specifically noted.
In this section we begin with models having  
$q = 0$ everywhere, so that cooling to low temperatures 
occurs in a small central region, 
then we investigate 
models in which $q$ is constant throughout the cooling flow.

\subsection{{\it Models Without Distributed Mass Dropout ($q =0$)}}

We begin by considering a perfectly homogeneous 
interstellar flow in which all the gas reaches 
the central computational zone ($r = 65$ pc) 
or its neighboring zones 
where it then cools to $T \ll 10^7$ K.
For comparison 
we discuss two cases: in Model 1 we ignore 
the gravity of this cooled gas; in Model 2 (and 
subsequent models) we include its gravitational influence 
on the flow.
The mass of gas that cools into the central gridzone 
is dynamically equivalent to a massive black hole, 
but we do not necessarily regard our 
calculation as a realistic model for black hole formation.
In Figure 2a we illustrate the radial variation of gas 
density and temperature for Models 1 and 2 
after the interstellar gas has evolved to $t = t_n = 13$ Gyrs.
Several global parameters for these and subsequent models
are listed in Table 1.

When the gravity of the cooled gas is considered (Model 2), 
the interstellar gas 
within a few kpc of the galactic center is compressed 
and sharply heated in the local potential. 
Such hot central thermal cores are not generally 
observed (but see Colbert \& Mushotzky 1998). 
The central mass in both models, 
$M_{cent} = 4.65 \times 10^{10}$ $M_{\odot}$ is  
about 16 times larger than the black hole mass found in NGC 4472, 
$M_{bh} = 2.9 \times 10^9$ $M_{\odot}$ (Magorrian et al. 1998). 
For these reasons neither Model 1 nor 2 
provides a realistic interstellar mass distribution for this 
galaxy. 

ROSAT band X-ray surface brightness profiles $\Sigma_x(R)$ 
for Models 1 and 2 
are shown in Figure 2b and the total ROSAT band luminosities 
are listed in Table 1.
For Model 1 the X-ray brightness peaks strongly 
in the galactic center, 
diverging from the observations within about 3 kpc; 
it was just this sort 
of disagreement that initially led to 
the assumption of distributed mass dropout in cooling flows
(Thomas 1986; 
Vedder, Trester, \& Canizares 1988;
Sarazin \& Ashe 1989).
If the efficiency of heating by early SNII were lowered, 
Models 1 and 2 could be made to agree better with 
$\Sigma_x$ observations in $r \gta 10$ kpc, 
but the computed $\Sigma_x$ would rise even further 
above the observations within a few kpc.
The X-ray luminosities $L_x$ 
for Models 1 and 2 listed in Table 1 are 
unreliable in part because of numerical inaccuracies caused 
by the extremely steep variation of gas temperature 
and density in the central 2 or 3 computational zones. 
This numerical difficulty is common to all cooling flows 
that proceed to the very center of the galaxy 
before cooling, for 
any reasonable central grid spacing. 
Since more $Pdv$ work is done on the flow in Model 2 
(where the gravity of cooled gas is included), we 
expect that $L_x$ should also be larger than for Model 1.
The opposite sense of the change in $L_x$ shown 
in Table 1 evidently results from numerical inaccuracies near 
the central singularity where zone to zone 
variations are no longer linear.
If the gas has not cooled before flowing 
into the core ($\lta 100$ pc), as in Models 1 and 2,
a significant fraction of the total X-ray emission 
should come from this very 
central region, again in disagreement with observations.

In Models 1 and 2 a sphere of cold ($T = 10$ K), dense gas
accumulates in the central gridzones and grows in mass and size over
the Hubble time.  
This sphere is an unrealistic artifact of our
computational assumptions. 
Therefore, to explore further the central
numerical difficulty with $L_x$ encountered in Models 1 and 2, we
considered two additional models using higher spatial resolution
(radius of central zone is only 15 pc), both of which include the
gravitational influence of cooled gas. 
The first model (Model 2.1) is 
similar to Model 2 with $q = 0$ in all zones.  
In the second model 
(Model 2.2) we set $q = 0$ in all zones except the central zone where
$q = 4$. 
Model 2.2 is the appropriate limit of the series
of models discussed below in which the cooling dropout is more and
more centrally concentrated.

While the flow in $r \gta 1$ kpc is very similar 
in Models 2.1 and 2.2, the behaviour at smaller radii 
is quite different. 
For example the flow parameters 
for Model 2.1 at 100 pc and $t_n = 13$ Gyrs are: 
$T = 7.9 \times 10^7$ K,
$n = 1$ cm$^{-3}$,
and 
$u = -320$ km s$^{-1}$. 
At the same radius and time 
for Model 2.2 the flow is quite different: 
$T = 6.3 \times 10^7$ K,
$n = 5$ cm$^{-3}$,
and 
$u = -60$ km s$^{-1}$.
At projected radius $R = 100$ pc, the ROSAT X-ray 
surface brightness in Model 2.2 is 20 times larger 
than Model 2.1 and the total ROSAT X-ray luminosity 
integrated over the entire cooling flow in Model 2.2 
is larger by a factor 32.
The differences between these two models result from 
upstream propagation of information 
about the central boundary conditions made possible 
by subsonic flow near the origin. 
Fortunately, these numerical and physical 
difficulties near the origin 
do not arise in more realistic cooling flows 
discussed below in which $q > 0$ at larger galactic radii.

\subsection{{\it True and Apparent Gas Density and Temperature}}

While solutions of the gasdynamical equations provide the
temperature as a function of physical radius $T(r)$, 
the observed temperature $T(R)$ is an
emission-weighted mean temperature 
along the line of sight at projected radius $R$.
For symmetric galaxies and at small galactic radii, these
two temperatures are nearly identical because of the
steep radial variation in X-ray emissivity.
The variation of 
temperature with physical radius $T(r,t_n)$ 
in the background flow is shown
with light solid lines in Figure 2a.

In the presence of spatially distributed 
cooling dropout, the local 
temperature is an emission-weighted mean of the 
background (uncooled) gas and the cooling regions.
Cooling is assumed to occur at a large number 
of cooling sites where the gas remains 
in pressure equilibrium as it cools, apparently unrestricted 
by magnetic stresses (see Mathews \& Brighenti 1999 for details).
The heavy solid lines in Figure 2a 
show the mean apparent temperature $T_{eff}(r,t_n)$
including contributions from locally cooling regions, 
\begin{equation}
T_{eff} = { T + q T \Delta_1(T) \over 1 + q \Delta_0 (T) }
\end{equation}
where $T$ is the background flow temperature and 
the slowly-varying functions $\Delta_i(T)$ are 
plotted in Brighenti \& Mathews (1998).
Note that the effective temperature is independent of the 
local gas density.
The temperature that is actually observed is an
average of $T_{eff}$ along the line
of sight; this temperature $T_{eff}(R,t_n)$
is shown with dotted lines in Figure 2a. 

Similarly, 
the apparent hot gas density is increased because of 
additional emission from denser, cooling-out gas.
The observed electron density shown in the Figure 2a 
is found by Abel inversion of 
the X-ray surface brightness distribution. 
When cooling sites are present, 
the local emissivity into the ROSAT energy band is
$$\varepsilon_{\Delta E} = (\rho_{eff}/m_p)^2 \Lambda_{\Delta E}(T)$$
$$= (\rho/m_p)^2 \Lambda_{\Delta E}(T)
[1 + q \Delta_0(T)] ~~~ {\rm ergs/sec~cm}^3.$$
The effective density is therefore
\begin{equation}
n_{eff} = n [ 1 + q \Delta_0(T)]^{1/2}
\end{equation}
where $n$ is the electron density of the background,
uncooled gas.
The observed (azimuthally-averaged)
densities in NGC 4472 plotted in Figure 2a 
should be compared with $n_{eff}(r,t_n)$ shown with 
heavy solid lines. 

\subsection{{\it Distributed Mass Dropout with Constant $q$}}

Lacking an acceptable physical model for 
spatially distributed mass dropout 
in galactic cooling flows, we are at liberty to choose 
any variation for the dropout coefficient $q(r)$, appropriately 
constrained {\it a posteriori} 
by the known dynamical mass from stellar velocities, 
the X-ray mass, the central black hole mass 
and the observed radial variation of hot gas density, 
temperature and X-ray surface brightness.
It is natural to begin with constant $q$ solutions 
similar to those 
considered by Sarazin \& Ashe (1989) in their steady state 
cooling flow solutions. 
For simplicity we assume that the mass of cooled gas
remains at the cooling site where it contributes to the
gravitational potential.

In the central panels of Figure 2a we 
illustrate the hot interstellar gas density and temperature that 
results after $t_n = 13$ Gyrs assuming uniform $q = 1$ 
and 4; 
these are 
listed in Table 1 as Models 3 and 4 respectively.
When $q$ is a constant independent of radius,  
the ratios of true to apparent 
values -- $n/n_{eff}$, $T/T_{eff}$ and $M_{tot}/M_x$ -- are 
also approximately uniform with galactic radius.
Here $M_{tot}(r)$ is the true total mass within $r$ and 
$M_x(r)$ is the value that would be determined from 
X-ray observations of the models 
by assuming hydrostatic equilibrium (as in Fig. 1).
The influence of constant-$q$ mass dropout 
is similar at all galactic radii since the 
factors that convert background temperature and density 
to effective values in equations (3) and (4) depend 
only on $T(r)$ which is slowly varying, not on $n(r)$.
Because of the enormous volume and time 
available to gas dropping 
out at large $r$, the total mass of cooled gas (listed in 
Table 1) becomes very large in the outer galaxy.
In Figure 3 we compare the radial distribution of 
stellar and dark halo mass with the 
distributed dropout mass that 
results after 13 Gyrs with $q = 1$ and 4.
The de Vaucouleurs ``stellar'' mass profile in Figure 3 
is constructed 
assuming uniform $M/L_B = 9.2$.

As $q$ increases from 1 to 4,
the background 
hot gas density (light solid lines in Fig. 2a) 
decreases and its radial gradient flattens, 
causing a rise in temperature to provide enough 
pressure to support the cooling flow atmosphere.
Note that the total dropped-out mass within 
$r_e/3$ decreases with larger $q$ (Figure 3 and Table 1).
Although more overall cooling dropout occurs as $q$ increases,
most of this cooling occurs at very large $r$ 
and, somewhat paradoxically, less gas remains for dropout 
closer to the galactic center, $r \lta r_e$.
However, the effective 
(i.e. apparent) density (heavy solid lines in Fig. 2a)
is less sensitive to $q$.
The $q = 1$ solution (Model 3) is preferred for its  
fit to observed temperatures while the $q = 4$ solution (Model 4) 
agrees better with the observed density in $r \lta 10$ kpc 
and almost exactly with the X-ray surface brightness 
(Fig. 2b).

In addition, the centrally concentrated 
dropped out mass in the $q = 1$ solution (Model 3) 
contributes 
a larger fraction of the dynamical mass in $r \lta r_e/3$   
and the total mass within the central gridzone
($r = 65$ pc) is almost equal to the known mass of the
black hole in NGC 4472 (Table 1). 
Evidently, cooled masses in 
models with $q < 1$ would exceed the central dark mass observed. 

For both values of $q$ considered, the dropped out mass 
listed in Table 1 
contributes appreciably to the total mass within $r = r_e/3$.
When both old stellar and dropout mass are included,
the $M/L_B$ values at $r_e/3$ shown in Table 1 -- 12.19 and 11.21 -- 
exceed both the dynamical value $M/L_B = 9.2$
found by van der Marel (1991) within $\sim 0.4r_e$ 
and the value $M/L_B(r_e/3) = 10.18$ 
in our models which includes a 
small additional contribution of non-baryonic dark matter. 
To quantify the influence of the dark baryonic dropout mass on 
the mass to light ratio we include in Table 1 an entry for 
$$\Delta_{m/l}(r_e/3) = 
{ (M/L_B)_{*,dh,do}(r_e/3) \over (M/L_B)_{*,dh}(r_e/3)} - 1$$
$$= {M_{*,dh,do}(r_e/3) \over M_{*,dh}(r_e/3)} - 1$$
where $M_{*,dh}(r_e/3) = 1.35 \times 10^{11}$ 
$M_{\odot}$ is the combined mass of 
luminous stars and dark halo matter at $r_e/3$ and 
$M_{*,dh,do}(r_e/3) = M_{tot}(r_e/3)$ also 
includes the dropout mass within this radius.
Since our computed total mass 
$M_{*,dh,do}(r_e/3)$ exceeds 
the observed dynamical value 9.2 (and also 10.18), 
to be fully consistent we should have chosen 
${M/L_B}_* < 9.2$ for the old stellar population
(see below), provided $M/L_B$ for stars produced from 
the cooled gas is infinite as we have assumed.

If these constant $q$ models are physically appropriate, 
the agreement between $M_x(r)$ and $M_*(r)$ 
in $(0.1 - 1)r_e$, as shown in Figure 1 
(and for NGC 4649), 
is surprising since $M_x(r_e/3)$ and $M_{*,dh}(r_e/3)$ 
differ by 10 - 40 percent (Table 1) for the constant $q$ models. 
This suggests that the mass to light ratio 
of the dropout stellar population is not infinite as 
we have assumed, but comparable with that of 
the old stellar population.

\section{MODELS WITH CENTRALLY CONCENTRATED DROPOUT}

We now seek evolutionary gasdynamical solutions with 
variable dropout coefficients $q(r)$ 
that strongly concentrate the mass dropout 
in the inner galaxy, $r \lta r_e$.
The limited spatial extent of optical emission lines 
in the cores of bright ellipticals suggests that cooling dropout 
occurs well within $r_e$.
For example, the H$\alpha$ + [NII] image of
NGC 4472 is observed out to approximately $\sim 0.25r_e$ or
2 kpc (Macchetto et al. 1996).
As hot interstellar gas cools in 
this region, its temperature pauses at $T \sim 10^4$ K where
the gas is heated and ionized by stellar UV radiation. 
Therefore, 
we consider parameters $r_{do}$ and $m$ in equation (2) 
that concentrate the mass dropout in the central region, 
but not at the very center as in Model 2. 

The dropout parameters $q_o$, $r_{do}$ and $m$,
are listed in Table 1 for Models 5, 6 and 7. 
The mass dropout in these three models is progressively 
more concentrated toward the galactic center.
The results for Model 5, for which the dropout scale length
is very large, $r_{do} = 15$ kpc, are similar in most 
respects to those of Model 4 except 
the massive dropout in the outer galaxy is no longer present.
In particular, 
the mass to light ratio at $r_e/3$ in Table 1 
is very similar for Models 4 and 5.

The current 
interstellar density and temperature variations for Model 6 
with $r_{do} = 2$ kpc are shown in Figure 2a 
and $\Sigma_x(R)$ is plotted in Figure 2b. 
The apparent density 
and $\Sigma_x$ (heavy solid lines) are considerably 
greater than observed values in $r \lta 3$ kpc,  
although the projected apparent temperature is a reasonable 
fit to the NGC 4472 data.
The radial distribution of 
dropout mass for Model 6 is shown in Figure 3.
From Table 1
the total (old stars, halo and dropout) mass to light ratio 
at $r_e/3$ is $M_{tot}/L_B = 13.03$, 
29 percent higher than van der Marel's value 
and 22 percent greater than the corresponding value for our 
background model galaxy. 

Also shown in Figures 2a and 2b are similar results for 
Model 7 in which the mass dropout, now 
approximated with a Gaussian, is concentrated within 
$r_{do} = 800$ pc. 
Model 7 is constructed so that 
most of the mass dropout occurs 
in $r \lta 0.1r_e$, corresponding to the region of apparent 
disagreement between $M_x$ and $M_*$ in Figure 1. 
It is of interest that the gas density and $\Sigma_x$ within 
1 kpc for Model 7, where the dropout is greatest, 
exceeds those of Model 2 in which there is no 
dropout at all. 
This can be understood from the flow velocity distribution.
For Model 2 without distributed dropout the inward moving 
gas velocity increases through the entire region 
illustrated and shocks at $r \ll 1$ kpc.
In the presence of dropout, the flow velocity 
at corresponding radii in Model 7 is much slower 
and reaches a maximum (negative) value 
near $r = 1$ kpc then approaches zero subsonically at the origin.
The density and $\Sigma_x$ 
enhancements in the background flow 
for Model 7 at $r \lta 3$ kpc 
are due to a local compression 
as the gas flows into slowly moving gas in the core. 
Within about 3 kpc, both the background and apparent 
densities exceed the observations by a larger factor than 
those of Model 6 
and $\Sigma_x$ also peaks 
unrealistically in this same region (Fig. 2b).
The dropout mass for Model 7 shown in Figure 3 equals 
that of the old population stellar mass 
at $r \approx 1$ kpc ($0.12 r_e$).
For these reasons Model 7 seems less satisfactory than 
Model 6, but neither is as generally successful as Model 3 
($q = 1$).

Although increased mass dropout can decrease the X-ray surface 
brightness $\Sigma_x$, as when uniform $q$ increased from 
1 to 4 (Models 3 and 4 in Fig. 2b), this is not always the case. 
When the dropout is concentrated more toward the galactic 
center, $\Sigma_x$ actually increases, as in the transition 
from Model 6 to 7.

It is interesting to determine the influence 
of distributed dropout on the total apparent 
mass $M_x(r)$ found from the model 
by assuming hydrostatic equilibrium (equation 1).
Due to the contribution of low temperature cooling regions 
to the total X-ray emission, $T(r)$ and 
therefore $M_x(r)$ is always lower 
than the true mass $M_{*,dh}(r)$ in cooling dropout regions. 
A difference in this sense is apparent in Table 1 at 
$r = r_e/3$ for Models 3 - 7; 
this is similar to the mass discrepancy  
in Figure 1 at $r \lta 0.1 r_e$.

The high central apparent gas density and surface brightness
for Model 6 shown in Figures 2a and 2b
are obvious problems for this model.
However, the gas density can be reduced
if a strong magnetic field 
or other non-thermal energy density 
is present in $r \lta 0.25r_e = 2$ kpc. 
Dynamically important magnetic fields
may also be required to fit the X-ray data
of NGC 4636 (Brighenti \& Mathews 1997a) and may be generally
expected in luminous ellipticals (Mathews \& Brighenti 1997;
Godon, Soker \& White 1998).
Additional non-thermal pressure support 
is also implied for Model 6 since
the effect of central cooling dropout fails to
lower the apparent mass $M_x(r)$ below the actual
mass $M_{tot}$ as much as the observed
deviation shown in Figure 1.
Like many bright ellipticals, NGC 4472 has a weak non-thermal
radio source within the central $\sim 4$ kpc,
indicating $B \sim 10 - 100$ $\mu$G (Ekers \& Kolanyi 1978). 

For all calculated models in Table 1 with distributed dropout, 
the total mass $M_{tot} = M_{*,dh,do}$ 
significantly exceeds the mass of the 
old stellar population plus dark halo $M_{*,dh}$ 
throughout the inner galaxy,
indicating that dropout material makes an important 
additional contribution to the total mass.
However, the apparent mass $M_x$ determined from equation (1) 
is less than $M_{tot}$ in the inner galaxy and, 
for Model 6, can also be less than $M_{*}$.
Of particular interest is the region 
$0.1 r_e \lta r \lta r_e$ 
(i.e., $-0.07 \lta \log r_{kpc} \lta 0.93$) in Figure 1. 
Although the agreement in Figure 1 is excellent in this 
region, for Models 6 and 7 the total mass is larger than 
$M_{*,dh}(r)$ and values of $\Delta_{m/l}$ in Table 1 
suggest that the dropped out 
mass contributes 25 - 35 percent of the total 
mass in this region.
Therefore, if the dropout mass is optically dark, 
the true stellar mass to light ratio of the 
old stellar population 
must be $M/L_B \approx 6$ rather than the value $M/L_B = 9.2$ 
found by van der Marel which includes the dropout mass. 

To investigate such a possibly more self-consistent 
old stellar component, 
we consider Model 8 based on the same $q(r)$ used in Model 6, but 
with $M/L_B = 6$ for the old stellar population.
For additional consistency 
in Model 8, $\alpha_*(t)$ is increased by the ratio of 
assumed stellar mass to light ratios 9.2/6 = 1.53 as described 
below; the total stellar mass ejected is identical
to that in Model 6. 
Figure 3 illustrates the dropout mass profile for Model 8. 
The apparent density, temperature and $\Sigma_x$ 
profiles for 
Model 8 shown in Figures 4a and 4b 
are very similar to those of Model 6, 
so this adjustment of the stellar $M/L_B$ has had 
little effect.

The radial mass profiles of Models 6 and 8 are 
compared in Figure 5. 
In this plot the open circles show the observed X-ray mass 
$M_x(r)$ for NGC 4472 using equation (1) and 
the solid lines shows $M_x(r)$ based on equation (1) 
using $n(r)$ and $T(r)$ from the models.
The dashed lines show $M_{tot}(r)$ and the dotted 
lines are the stellar mass $M_*(r)$ based on 
a de Vaucouleurs profile with $M/L_B = 9.2$ in the 
upper panel (Model 6) and $M/L_B = 6$ in the lower panel 
(Model 8).
The superiority of Model 8 is evident from the 
closer agreement between the 
X-ray mass data points for NGC 4472 and 
the solid line for that model.
This agreement for Model 8 would be even closer
in the range $\log r \approx 0.5 - 1$ if we had 
used a dark halo mass profile less centrally 
peaked than NFW.
Model 8 may provide the most self-consistent 
overall fit to the mass
constraints for NGC 4472; 
if so, the old stars in NGC 4472 
have a mass to light ratio $M/L_B \approx 6$, 
about $\sim 30$ percent lower than the mass to light 
ratio determined from stellar dynamics.

In summary, none of our  
models is fully satisfactory in every respect. 
The total mass to light ratio 
(including dropout mass)  
in $r \lta r_e/3$ is 10 - 35 percent higher than 
the value for the underlying galaxy.
However, in Model 8 in which $M/L_B = 6$ for the 
old stars, the difference between the
X-ray mass determined from the models and NGC 4472 
is appreciably reduced.
Nevertheless, 
the central apparent gas density 
and X-ray surface brightness in Model 8
are still larger than observed, requiring additional
non-thermal support.
There is no independent theoretical justification 
for the dropout profile $q(r)$ assumed in 
Models 6 and 8; 
as explained earlier, the dropout distribution depends on 
unknown interstellar entropy fluctuations. 
Nevertheless, for all models considered here 
the mass of cooled interstellar gas contributes 
significantly to the total mass 
and the dynamically determined mass to light ratio 
within the inner galaxy. 

\section{FINAL REMARKS AND CONCLUSIONS}

In this series of calculations 
we have taken a census of all baryons involved in 
the evolution of a large elliptical galaxy:
the original stellar component,
the interstellar medium, and -- of most interest -- the 
small but troublesome mass of hot gas 
that cools over cosmic time.
We have shown that 
the radial distribution of cooled interstellar gas 
influences dynamical and X-ray determinations of 
the total interior mass and the radial profiles 
of apparent density, temperature and X-ray brightness of 
the hot gas.
In our models, cooled gas is 
slowly deposited in the central galaxy 
$r \lta r_e$ (see Figure 3) 
as indicated by the extent of observed H$\alpha$ emission.
Since there is little or no direct observational evidence for 
the mass that has dropped out in ellipticals like NGC 4472, 
the cooled mass must either 
be dark at optical and radio frequencies 
or indistinguishable from the old stars.
Low mass stars are an obvious and physically reasonable  
endstate for the cooled gas 
(Ferland, Fabian \& Johnstone 1994; 
Mathews \& Brighenti 1999).
We also suppose that 
the cooled gas remains at the dropout site 
where it contributes to the galactic potential.

\subsection{{\it Reducing Stellar Mass Loss}}

In an attempt to reduce the influence of cooling and cooled 
gas on the models, we have altered many of the 
model parameters.
The chosen cosmology ($\Omega = 1$; $\Omega = 0.3$ 
and $\Omega_{\Lambda} = 0.7$, etc.) or 
the baryon fraction $\Omega_b$ have little 
influence on the total dropout mass.
Changing the times when the stars and 
the galactic potential form ($t_{*s}$ and $t_{*}$) or the 
spatial scale of the release of SNII energy 
within reasonable limits
have only a modest influence on the dropout mass 
that accumulates by time $t_n = 13$ Gyrs. 
Increasing the interval $t_{*s} - t_*$ between star and 
galaxy formation does reduce the total dropout mass, 
but this interval cannot be too large  
since luminous ellipticals are observed at large redshifts.
The cooling dropout would not be dramatically reduced  
if massive ellipticals were all only 
a few Gyrs old since most of the stellar mass loss 
occurs just after star formation; however, many or 
most luminous ellipticals are thought to be very old. 

Perhaps the most effective way to preserve the excellent
agreement between $M_{*}$ and $M_x$ in Figure 1
in $0.1r_e \lta r \lta 1 r_e$, without constraining
the mass dropout profile $q(r)$, would be to reduce
the total mass that has cooled over cosmic time.
The most sensitive parameter influencing the cooled mass is 
the specific rate of stellar mass loss, $\alpha_*(t)$. 
Although the total mass of hot gas increases with 
time due to the continued influx of secondary infalling gas, 
the mass of hot gas within the optical galaxy 
originates mostly from stellar mass loss and 
the X-ray luminosity there scales as 
$L_x \propto \alpha_*(t) \propto t^{-1.3}$ 
(Appendix B of Tsai \& Mathews 1995).
Computed models similar to those described here 
but with arbitrarily reduced $\alpha_*(t)$ 
fit the $n(r)$ and $\Sigma_x(R)$ data rather well 
at time $t_n$ and produce much less cooling dropout. 
However, $\alpha_*(t)$ cannot be lowered 
without also increasing the stellar mass.
For all reasonable power law initial mass functions, 
$\alpha_*(t)$ varies inversely  
with the stellar mass to light ratio:
$$\alpha_*(t) \equiv { d M_* /dt \over M_*}
= \left[{ d M_* /dt \over L_B}\right] {1 \over (M_*/L_B)}
\propto {1 \over (M_*/L_B)}.$$
The physical explanation for the constancy 
of $(d M_* /dt)/L_B$  is that both $L_B$ (dominated by
post-main sequence stars) and $d M_* /dt$ depend on
the instantaneous rate that stars leave 
the main sequence so this IMF-dependent factor 
cancels out (e.g. Renzini \& Buzzoni 1986).
To illustrate this result,
we use the Renzini-Buzzoni procedure 
and plot in Figure 6 the relationship between 
$\alpha_*$ and $M/L_B$ at time $t_n = 13$ Gyrs for 84 
power law IMFs with slopes $x = $0.6, 0.8, 1.0, 1.2, 1.4, 1.6,
and 1.8, each with lower and upper mass limits of 
$m_{\ell} = $0.01, 0.032, 0.1, 0.32
and $m_u = $10, 32, 100.
The remarkable linearity in Figure 6 shows 
that the quantity in square brackets in the equation above 
is almost invariant to large  
changes in the slope or mass limits for power law IMFs.

Therefore, if $\alpha_*(t)$ is reduced by 2 or 3 in an attempt 
to reduce the total dropout mass, 
the stellar mass (and $M/L_B$) must be increased by the same 
factor; as a result the total mass ejected, 
$\int \alpha_* M_* dt$, and the total dropout mass 
do not change. 

\subsection{{\it Effect of Galactic Rotation}}

Although massive ellipticals are not rotationally
flattened, they do rotate significantly, 
e.g. $(v/\sigma)_* = 0.43$ for NGC 4472 
(Faber et al. 1997).
If the hot interstellar gas rotates
in the same sense as the bulk of mass-losing stars, 
stars formed from the cooled gas
should form into a disk of scale $\sim r_e$
(Brighenti \& Mathews 1997b), 
although the development of such disks is likely to be 
suppressed by the mass dropout process. 
Nevertheless, to the extent that the cooled gas has a disk-like 
distribution,  
its global influence on the stellar
dynamics in $r \lta 0.4r_e$ would be less than if the
same dark mass were distributed spherically.
Remarkably, there is  
no observational evidence at present for
rotational flattening in the X-ray images of giant
ellipticals like NGC 4472.

\subsection{{\it Non-Baryonic Dark Matter within $r_e$}}

Dynamical determinations of the mass to light ratio from 
stellar velocities reflect the entire mass within 
the stellar orbits, including non-baryonic mass.
The NFW halo we use in our models for NGC 4472 agrees with 
X-ray observations in the extended halo but is 
slightly too massive 
near $r \sim r_e$ relative to the X-ray mass 
$M_x(r)$ observationally determined for NGC 4472. 
This may indicate that dark halos are less 
centrally peaked than NFW (see Kravtsov et al. 1998).
The mass of our NFW halo model contributes about 10 percent to 
the dynamical mass to light ratio measured within $r_e/3$,
but the NFW profile is probably disturbed in this region. 
When the dominant baryonic mass in $r \lta r_e$ compressed 
to form the de Vaucouleurs profile, we expect that the NFW 
halo was dragged inward and distorted. 
However, the dark halo cores of the earlier 
galactic condensations 
that merged to form the elliptical may have expanded 
due to starburst driven galactic winds. 
Because of these various counteracting effects, 
the small non-baryonic contribution 
to dynamical mass to light determinations is uncertain.

\subsection{{\it Contribution of Dropout to ``Stellar'' $M/L$}}

For all models studied here -- based on a wide variety 
of mass dropout profiles $q(r)$ -- 
the mass of cooled interstellar 
gas contributes substantially to the 
total mass within $\sim 0.4r_e$ where 
the stellar mass to light ratio is determined from 
stellar velocities.
If optically dark low mass stars form from the cooled gas, 
the ``stellar'' mass to light ratios in the literature 
refer to two distinct 
stellar populations having radically different initial 
mass functions and spatial distributions. 
The stellar mass to light ratio of the original, optically luminous 
stellar population is lower than published values indicate. 

A complete solution of this problem 
requires a better understanding of 
the physics of star formation and the processes that 
control the stellar IMF. 
We have assumed here that a Salpeter 
IMF provides a satisfactory approximation to the 
original single-burst star formation at early times.
Yet we argue that the younger dropout IMF is strongly 
skewed toward low mass stars. 
The IMF has evolved over time. 
It is possible therefore that the early, more 
intense mass dropout resulted in 
a more nearly Salpeter-like IMF, 
producing a fraction of currently observed 
luminous stars in ellipticals. 
If so, this early dropout would not contribute to the 
excess dropout mass that we find in our models, 
but would have already been included in the 
original de Vaucouleurs population.
High density, metal enriched 
stellar cores in elliptical galaxies may have derived from 
normal-IMF star formation from early, more intense 
interstellar cooling. 
This is similar to assumptions made for 
dissipative galactic core formation 
from the convergence of 
gas following major mergers (Mihos \& Hernquist 1996).
While such notions cannot be entirely dismissed,
the approximate universality of the de Vaucouleurs 
light profile 
among ellipticals may argue against a dual formation process 
for the radial distribution of luminous stars: 
violent relaxation and cooling flow dropout.

Throughout this discussion we have assumed that the stellar 
population formed from cooling flow dropout is optically dark. 
Although there is little or no evidence that normal massive 
OB stars (or SNII) are present in elliptical galaxies, 
it is possible that younger stars having masses up to 
$\sim 1 - 2$ $M_{\odot}$ are present and that 
such intermediate mass stars could form from the 
cooled gas (Mathews \& Brighenti 1999).
This type of dropout stellar population 
could contribute to the total optical light. 
If the mass to light ratio of dropout and old stellar 
populations are similar, the dropout 
component could be difficult to detect by the means 
we have discussed here and its perturbation on 
the observed $M/L_B$ would be greatly lessened.
In this case the dropout population would introduce 
an additional radial light profile that would differ 
slightly from that of the old stellar population. 
The dropout mass profiles in Figure 3 indicate that 
Models 4 and 6 would be rather difficult to detect 
against the background stellar light.
Intermediate mass dropout stars could therefore 
provide a satisfactory resolution to the problems 
we have discussed here.

\subsection{{\it Influence of Dropout on the Fundamental Plane}}

The ensemble of elliptical galaxies is known to have 
global parameters that deviate slightly from the assumptions 
of virial equilibrium and homologous structure.
The deviation of this fundamental plane relationship 
is in the sense that the dynamical mass to 
light ratio increases with galactic mass,
$M/L_B \propto M^{0.24}$
(Dressler et al. 1987; Djorgovski \& Davis 1987).
Such a non-homologous deviation could in principle be produced 
by the small amount of cooled interstellar gas $M_{cg}$ 
provided it increases appropriately with $M_*$. 
To test this idea, we performed an identical hydrodynamic 
calculation for an elliptical galaxy having a mass one fourth 
that of NGC 4472. 
The dark halo mass was also reduced by the same factor 
but the cosmological environment 
and mass dropout distribution 
were identical to those used for NGC 4472, 
scaled to a smaller $r_e$.
We found that the amount of mass dropout 
$M_{cg}$ is {\it higher} in smaller ellipticals 
relative to the total baryonic mass $M_*$. 
This is opposite to the trend observed in the 
fundamental plane.
However, if hot gas and dark matter in 
the outer halos of smaller ellipticals 
is tidally stripped in group environments, 
as suggested by Mathews \& Brighenti (1998b), 
then the mass of cooled gas would be reduced and 
its effect on the fundamental plane would be reduced. 
Nevertheless, explanations of the deviations of 
the fundamental plane from virial scaling 
must recognize the possible 
additional influence of dropout 
mass, regardless of the trend of $M_{cg}/M_*$ with 
$M_*$.

\subsection{{\it Conclusions}}

Using simple spherical gas-dynamical
models for the evolution of interstellar gas  
and data from the well-observed elliptical
NGC 4472, we reach the following conclusions:

\noindent
(1) If the hot interstellar 
gas cools only in the very center of NGC 4472
for $\sim 10$ Gyrs, the total accumulated mass there 
would be $\gta 10$ times larger than the 
mass of the central black hole
observed. If such large
concentrated masses were generally 
present in bright ellipticals,
interstellar gas in $r \lta 0.1r_e$ would be 
compressed and heated to temperatures $> 1$ keV. 
Such hot thermal cores are not generally observed.
We conclude that 
the cooling dropout in massive 
ellipticals must occur before the gas reaches
the galactic center.
The hypotheses of distributed 
mass dropout and low mass star formation 
were proposed many years ago 
(Fabian, Nulsen \& Canizares 1982; 
Thomas 1986; 
Cowie \& Binney 1988;
Vedder, Trester, \& Canizares 1988;
Sarazin \& Ashe 1989;
Ferland, Fabian, \& Johnstone 1994).
However, these historical arguments were 
generally based on the notion that mass dropout would 
help reduce computed X-ray surface brightness 
profiles at small projected radii, as required 
by the observations, but we have 
shown here that in some cases enhanced dropout 
at small galactic radii can cause 
$\Sigma_x$ to increase, not decrease. 
(The total bolometric X-ray luminosity $L_x$ 
should always be lower in distributed cooling models since the 
hot gas experiences only a fraction of the galactic 
potential.) 
The best arguments for distributed cooling dropout 
are (i) limits on the central black hole mass and (ii) the 
absence of rotational flattening in X-ray images.

\noindent
(2) We have considered a wide variety of possible 
mass profiles for the radial deposition of 
cooled interstellar gas in NGC 4472. 
The dropout mass is assumed to be optically dark, 
consistent with the formation of very low mass stars. 
In every case the (stellar plus dropout)
mass to light ratio at $\sim r_e/3$ significantly exceeds 
the mass to light ratio determined from stellar velocities. 
If the dropout mass is optically dark, dynamical 
mass to light ratios in luminous ellipticals 
should be substantially enhanced by 
dark baryonic matter. 
In this case 
the true $M/L_B$ for luminous stars may be $\sim 30$ percent 
smaller than published values. 

\noindent
(3) The excellent agreement
between the X-ray and ``stellar'' mass in NGC 4472
shown in Figure 1 (and also NGC 4649) 
in the range $0.1r_e \lta r \lta 1 r_e$
may be a coincidence if our estimates 
of the cooling flow dropout mass are correct 
and if this mass is non-luminous.

\noindent
(4) Dynamical mass to light determinations within $r_e$ refer 
to a superposition of two stellar populations: an old luminous 
population with a de Vaucouleurs galactic profile and 
a younger population having a bottom-heavy IMF and 
a different galactic mass profile.
If the younger population is optically dark,
the mass to light ratio is not likely to be
constant with galactic radius within $r_e$.

\noindent
(5) It is not possible to reduce the total amount 
of mass deposited from the cooling flow simply by lowering 
the specific rate of stellar mass loss
$\alpha_*(t) \equiv (dM_*/dt)/M_*$.
For all reasonable power law initial mass functions 
we show that $\alpha_* \propto (M_*/L_B)^{-1}$.
For given $L_B$, larger stellar masses $M_*(r)$ 
must accompany lower values
of $\alpha_*$ so the total amount of mass
ejected from stars $\propto \int \alpha_*(t)M_{*} dt$
is nearly independent of the IMF.

\noindent
(6) Among the models we consider, those with 
centrally concentrated mass dropout perform best in
minimizing the overall disagreement with the central $M/L_B$
and the X-ray determined mass in $0.1r_e \lta r \lta 1 r_e$.
Constant $q$ 
models in which the dropout is proportional to the
local gas emissivity at every radius deposit less mass
in $0.1r_e \lta r \lta 1 r_e$, but may have gas temperatures
that are too low ($q \gta 4$) 
or central masses that are too large ($q \lta 1$).

\noindent
(7) Even in the presence of mass dropout, 
the computed central interstellar gas 
densities and X-ray surface brightnesses $\Sigma_x(R)$ are 
generally too large. 
Such deviations would be reduced if the hot gas 
is partially supported in $r \lta 0.1r_e$ by 
magnetic or other non-thermal pressure 
associated with the extended radio source. 

\noindent
(8) If the stellar population formed from cooled interstellar 
gas extends to intermediate masses, 
$\sim 1 - 2$ $M_{\odot}$, its mass to light ratio may blend 
with that of the older population. 
In this case the dropout mass would already be represented 
in the de Vaucouleurs profile representing the 
stellar mass distribution in our models.
Dynamical determinations of 
$M/L_B$ would be a weighted mean of the two 
populations. 
If the dropout stellar population is luminous, 
some of the difficulties we have discussed here would 
be alleviated, but not those regarding $\Sigma_x(R)$.

\noindent
(9) If the influence of rotation and non-thermal pressure 
can be understood, 
high resolution images of the central regions of 
elliptical galaxies using the {\it Chandra} (AXAF) satellite 
may detect the presence of dark baryonic dropout material 
and an accurate determination of the mass to light ratio 
of the old stellar population.

\acknowledgments

Thanks to Karl Gebhardt for providing useful
information.
Studies of the evolution of hot gas in elliptical galaxies 
at UC Santa Cruz are supported by
NASA grant NAG 5-3060 and NSF grant AST-9802994  
for which we are very grateful. 
FB is supported
in part by Grant MURST-Cofin 98.







\clearpage


\begin{deluxetable}{crccccccccccc}
\tabletypesize{\scriptsize}
\tablewidth{60pc}
\tablenum{1}
\tablecolumns{13}
\tablecaption{MASS DROPOUT MODELS FOR NGC 4472\tablenotemark{a}}
\tablehead{
\colhead{Model} &
\colhead{$q_o$} &
\colhead{$r_{do}$} &
\colhead{$m$} &
\colhead{$M_{cg}$\tablenotemark{b}} &
\colhead{$M_{cg}({r_e \over 3})$\tablenotemark{c}} &
\colhead{$M_{tot}({r_e \over 3})$\tablenotemark{d}} &
\colhead{$M_x({r_e \over 3})$\tablenotemark{e}} &
\colhead{${M_{tot}\over L_B}({r_e \over 3})\tablenotemark{f}$} &
\colhead{$\Delta_{m/l}({r_e \over 3})\tablenotemark{g}$} &
\colhead{$M_{cent}$\tablenotemark{h}} &
\colhead{$L_{x,bck}^{ros}$\tablenotemark{m}} &
\colhead{$L_{x,tot}^{ros}$\tablenotemark{n}} \cr
\colhead{} &
\colhead{} &
\colhead{(kpc)} &
\colhead{} &
\colhead{($10^{10} \; M_{\odot}$)} &
\colhead{($10^{10} \; M_{\odot}$)} &
\colhead{($10^{11} \; M_{\odot}$)} &
\colhead{($10^{11} \; M_{\odot}$)} &
\colhead{($M_{\odot}/L_{B \odot}$)} &
\colhead{} &
\colhead{($10^{10} \; M_{\odot}$)} &
\colhead{($10^{40}$~erg~s$^{-1}$)} &
\colhead{($10^{40}$~erg~s$^{-1}$)} \cr
}
\startdata
1 & 0  & ... & ... & $4.64$ & 4.64 & $1.35\tablenotemark{k} $
& $1.36 $ & $13.70$ & 0.00 & $4.64 $ & 28.6 & 28.6 \cr
2  &  0  & ... & ... & $4.66$ & 4.66 & $1.81 $
& $1.70$ & $ 13.71$ & 0.34 & $4.66 $ & 21.7 & 21.7 \cr
3 & 1 & ... & ... & $36.1 $\tablenotemark{i}
& $2.64 $ &
$1.61 $ & $1.26 $ & $12.19 $& 0.19 & $0.254$ &
18.7 & 36.4 \cr
4 & 4 & ... & ... & $81.1 $\tablenotemark{j}
& $1.36$ &
$1.48 $ & $0.97 $ & $11.21$ & 0.10 & $ 0.013$ &
8.8 & 42.6 \cr
5 & 4 & 15 & 1 & $4.65 $ & $1.55$
& $1.50 $ &
$1.01 $ & $11.36$ & 0.11 & $ 0.013$ & 13.9 & 27.7 \cr
6 & 4 & 2 & 1 & $4.65 $ & $3.76$ &
$1.72 $ & $1.51 $ & $13.03$ & 0.27 & $ 0.015 $ &
22.1 & 33.5 \cr
7 & 4 & 0.8 & 2 & $4.65 $ & $4.65$ &
$1.81$ & $1.76$ & $13.71$ & 0.34 & $ 0.017$ &
29.1 & 43.3 \cr
8\tablenotemark{l} & 4 & 2 & 1 & $4.60 $ & $3.65$ &
$1.29 $ & $1.17 $ & $9.74$ & 0.40 & $ 0.012 $ &
20.0 & 28.1 \cr
\enddata
\tablenotetext{a}{All masses evaluated at time $t_n = 13$ Gyrs.}
\tablenotetext{b}{Total cooled mass within 1 Mpc.}
\tablenotetext{c}{Mass of cooled gas within ${r_e/3} = 2.86$ kpc.}
\tablenotetext{d}{Total mass within ${r_e/3}$; $M_{*,dh}(r_3/3) = 1.36
\times 10^{11}$ $M_{\odot}$.}
\tablenotetext{e}{Mass of cooled gas at ${r_e/3}$ evaluated using
hydrostatic equilibrium.}
\tablenotetext{f}{$M{*,dh}/L_B = 10.18$ 
is the value for the stars and dark halo for Models 1-7.}
\tablenotetext{g}{Relative contribution of dropout mass 
to $M/L_B$ at $r_e/3$.}
\tablenotetext{h}{Mass cooled into central gridzone;
the central black hole in NGC 4472 has mass
$M_{bh} = 0.29 \times 10^{10}$ $M_{\odot}$.}
\tablenotetext{i}{Mass within 100 kpc is $4.72 \times 10^{10}$
$M_{\odot}$}
\tablenotetext{j}{Mass within 100 kpc is $4.96 \times 10^{10}$
$M_{\odot}$}
\tablenotetext{k}{Mass of cooled gas is not included.}
\tablenotetext{l}{With lower $M/L_B$ for old stars.}
\tablenotetext{m}{ROSAT X-ray luminosity (0.2 - 2 keV) of 
background flow}
\tablenotetext{n}{Total ROSAT X-ray luminosity including 
emission from dropout}
\end{deluxetable}

\normalsize


\begin{deluxetable}{ccc}
\scriptsize
\tablewidth{15pc}

\tablenum{1}
\tablecolumns{3}
\tablecaption{MASS DROPOUT MODELS FOR NGC 4472 -- TABLE 1 CONTINUED
HERE BECAUSE OF AASTEX BUG}
\tablehead{
\colhead{Model} &
\colhead{$L_{x,bck}^{ros}$\tablenotemark{m}} &
\colhead{$L_{x,tot}^{ros}$\tablenotemark{n}} \cr
\colhead{} &
\colhead{($10^{40}$~erg~s$^{-1}$)} &
\colhead{($10^{40}$~erg~s$^{-1}$)} \cr
}
\startdata
1 &  28.6 & 28.6 \cr
2 & 21.7 & 21.7 \cr
3 & 18.7 & 36.4 \cr
4 & 8.8 & 42.6 \cr
5 & 13.9 & 27.7 \cr
6 & 22.1 & 33.5 \cr
7 & 29.1 & 43.3 \cr
8 & 20.0 & 28.1 \cr
\enddata
\end{deluxetable}

\normalsize


\clearpage


\vskip.1in
\figcaption[aasdropoutfig1.ps]{
(a) Interstellar gas temperature and density in NGC 4472. 
top: fit to ROSAT data for $T(R)$; bottom: fit to ROSAT
data for $n(r)$ (open circles) and {\it Einstein} data (closed circles);
(b) {\it solid curve}: $M_x(r)$ determined from above data 
using equation (1); 
{\it long-dashed curve}: de Vaucouleurs 
mass profile $M_*(r)$ assuming $M_*/L_B = 9.2$.
\label{fig1}}

\vskip.1in
\figcaption[aasdropoutfig2.ps]{
(a) Comparison of observed hot gas density and temperature 
in NGC 4472 with various hydrodynamical models at time 
$t_n = 13$ Gyrs.
The source of observed data in all panels is identical to 
that in Figure 1.
{\it Light solid lines:} density and temperature of the 
background flow as functions of physical radius $r$;
{\it heavy solid lines:} apparent density and temperature 
as functions of $r$ including emission from cooling regions;
{\it dotted lines:} apparent gas temperature 
as function of projected 
radius, similar to the temperature data points.
(b) Comparison of ROSAT and {\it Einstein} 
X-ray surface brightness distributions in NGC 4472 
with various models. 
{\it Light solid lines:} emission just from the 
background cooling flow; 
{\it heavy solid lines:} combined emission from 
background and distributed cooling regions.
\label{fig2}}

\vskip.1in
\figcaption[aasdropoutfig3.ps]{
Baryonic and non-baryonic mass distributions in NGC 4472.
{\it Light solid line:} de Vaucouleurs profile of 
dynamical mass based on $M/L_B = 9.2$;
{\it heavy solid line:} NFW dark halo profile for best fit 
to NGC 4472 data;
{\it short dashed line:} cooled dropout mass $M_{cg}$ for 
Model 1;
{\it long dashed line:} cooled dropout mass $M_{cg}$ for
Model 4;
{\it dotted line:} cooled dropout mass $M_{cg}$ for
Model 6;
{\it short dash-dotted line:} cooled dropout mass $M_{cg}$ for
Model 7;
{\it long dash-dotted line:} cooled dropout mass $M_{cg}$ for
Model 8.
\label{fig3}}

\vskip.1in
\figcaption[aasdropoutfig4.ps]{
(a) Comparison of observed hot gas density and temperature
in NGC 4472 with Model 8 at time
$t_n = 13$ Gyrs. 
Data and curve designations are identical to those in 
Figure 2a.
(b) X-ray surface brightness profiles at $t_n = 13$ Gyrs 
for Model 8.
Data and curve designations are identical to those in
Figure 2b.
\label{fig4}}

\vskip.1in
\figcaption[aasdropoutfig5.ps]{
Mass profiles at $t_n = 13$ Gyrs for Model 6 (upper panel) 
and Model 8 (lower panel).
{\it Solid lines:} X-ray mass profiles $M_x(r)$
derived from the computed models using equation (1);
{\it dashed lines:} actual mass profiles $M_{tot}(r)$
for each model;
{\it dotted curves:} de Vaucouleurs profiles $M_*(r)$
based on uniform mass to light ratios, 
$M/L_B = 9.2$ for Model 6 and $M/L_B = 6$ for Model 8.
{\it Circles:} data points show the 
apparent mass distribution 
$M_x(r)$ using NGC 4472 observations and equation (1).
\label{fig5}}

\vskip.1in
\figcaption[aasdropoutfig6.ps]{
Plot of the specific mass loss rate $\alpha_*(t_n)$ (in sec$^{-1}$)
against the mass to light ratio $M_*/L_B$ in solar units 
for 84 power law initial mass functions with varying slopes and 
mass cutoffs as described in the text.
\label{fig6}}

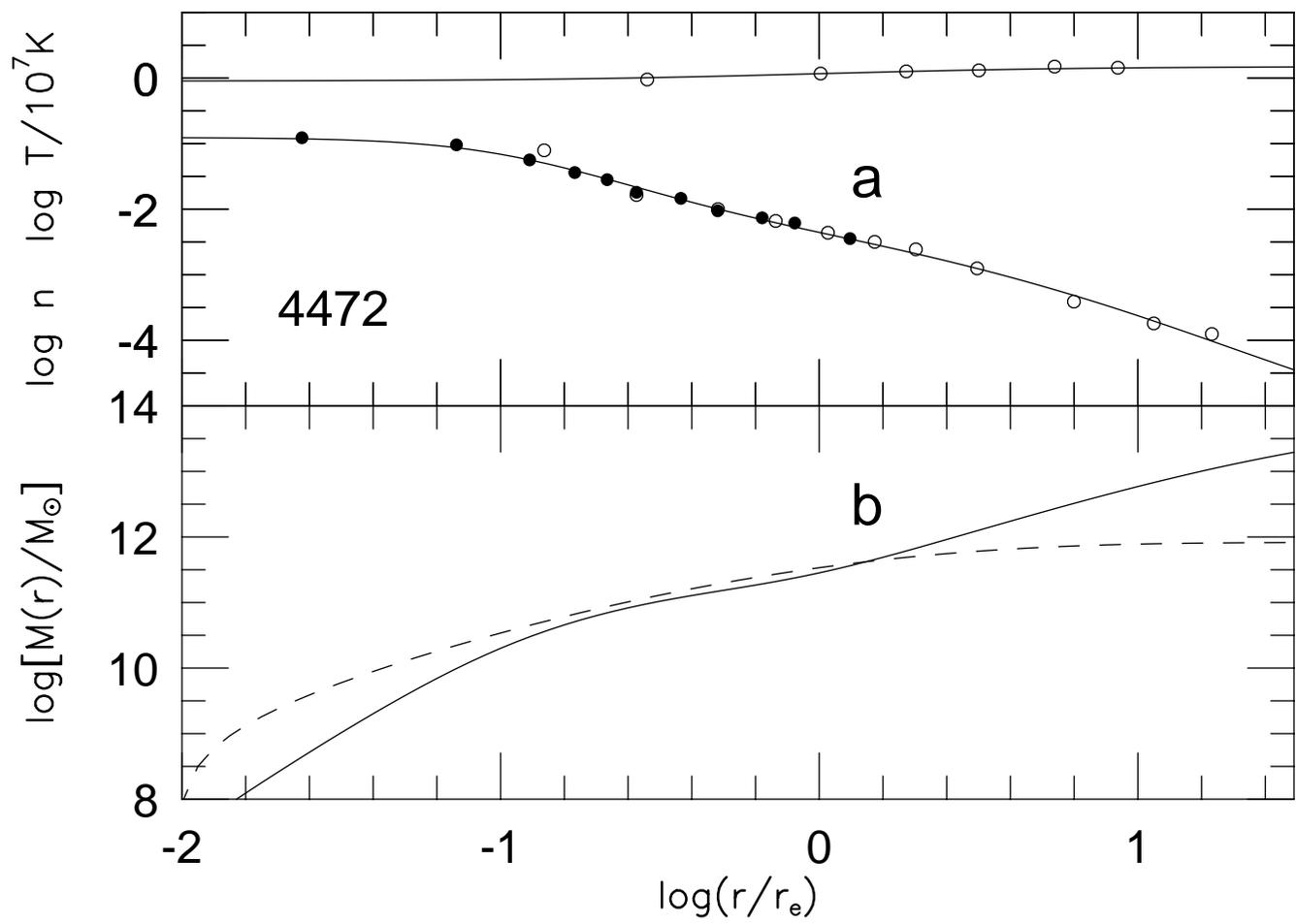
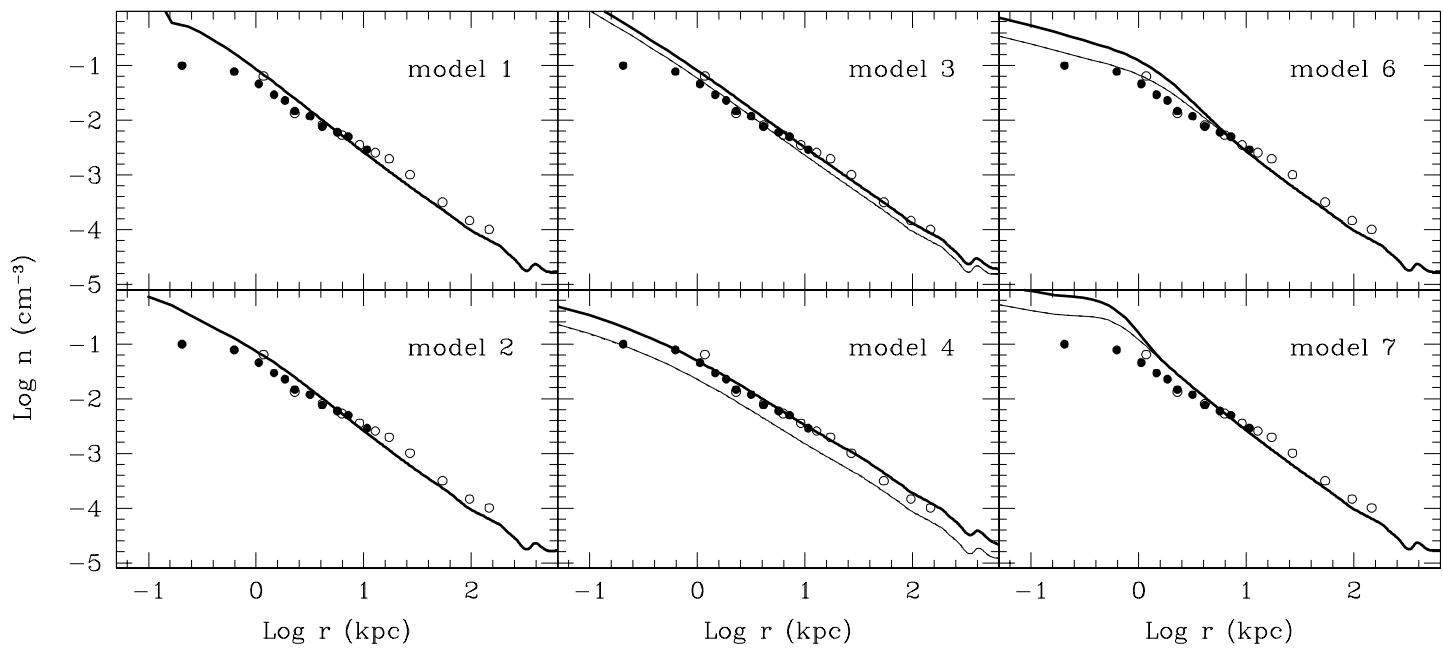
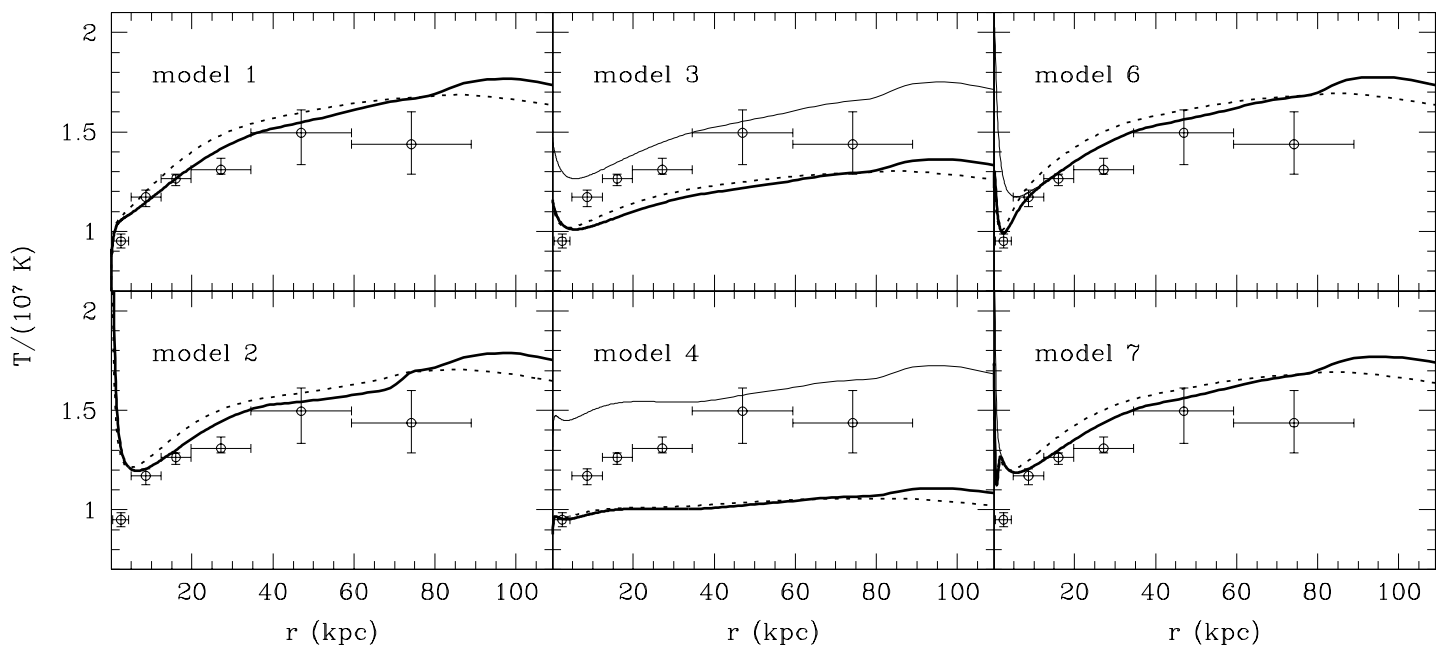
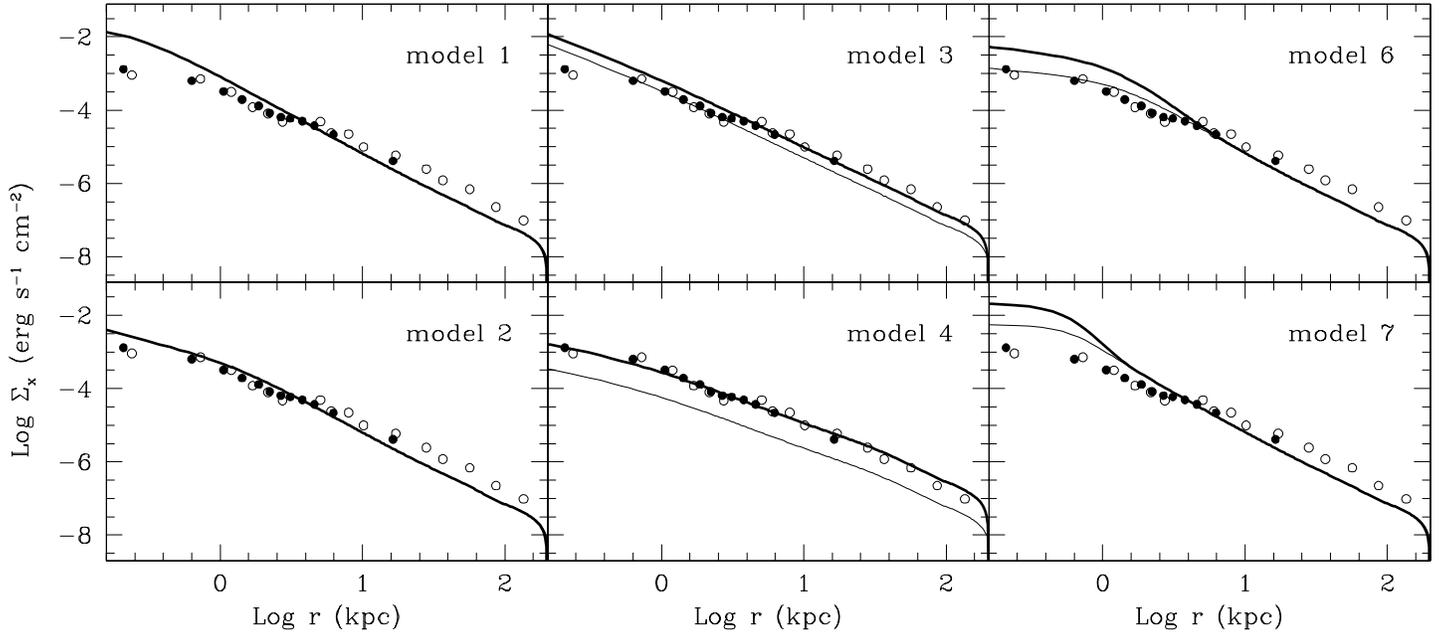
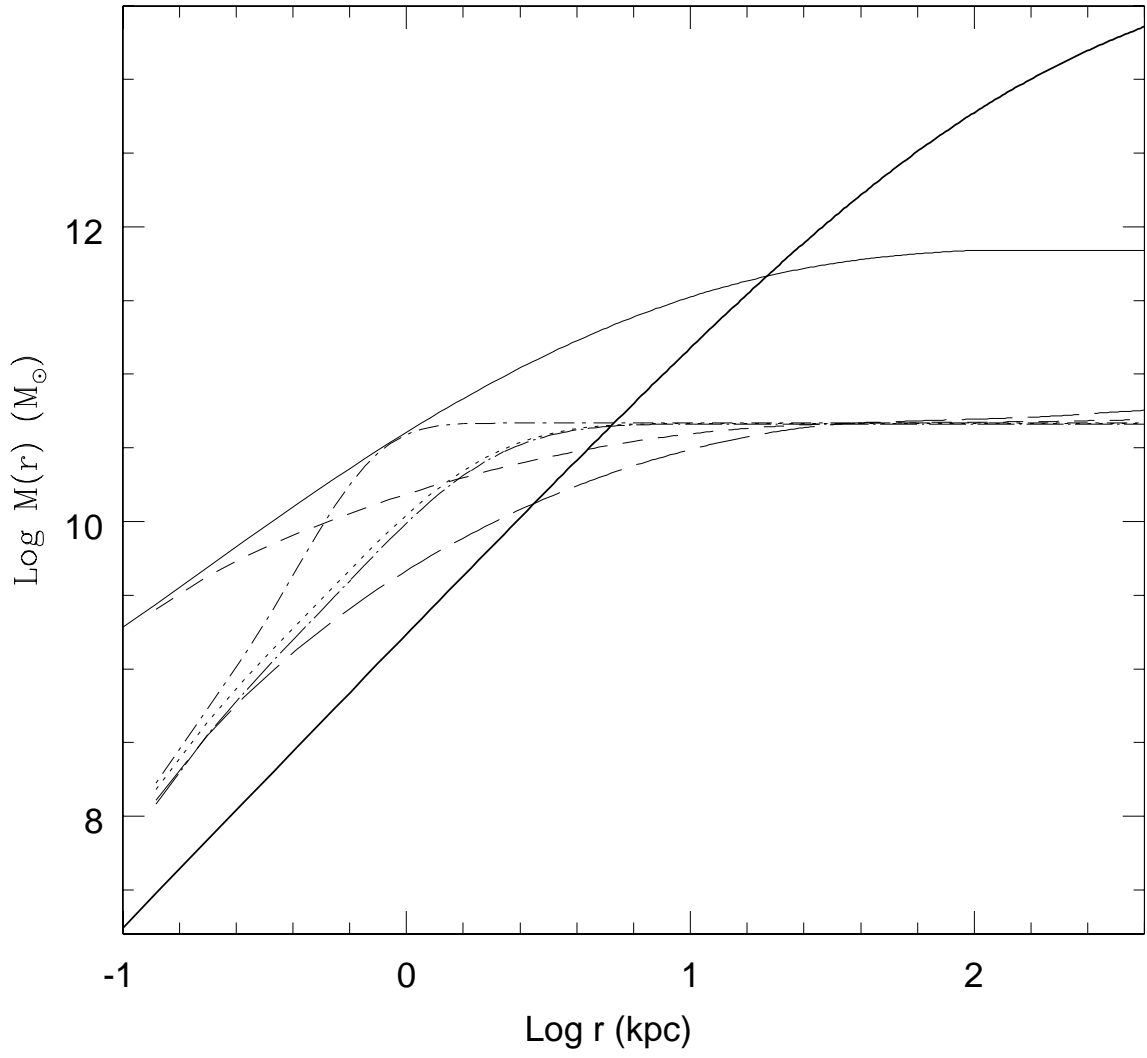
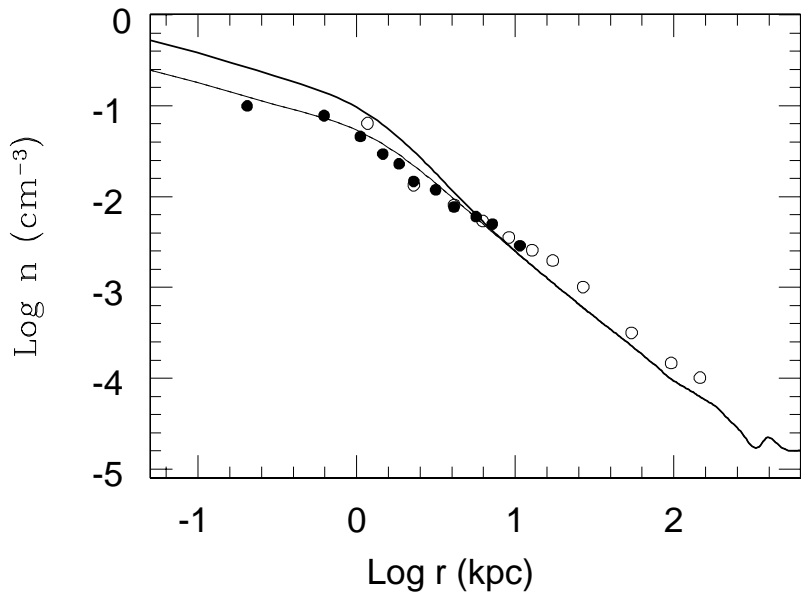
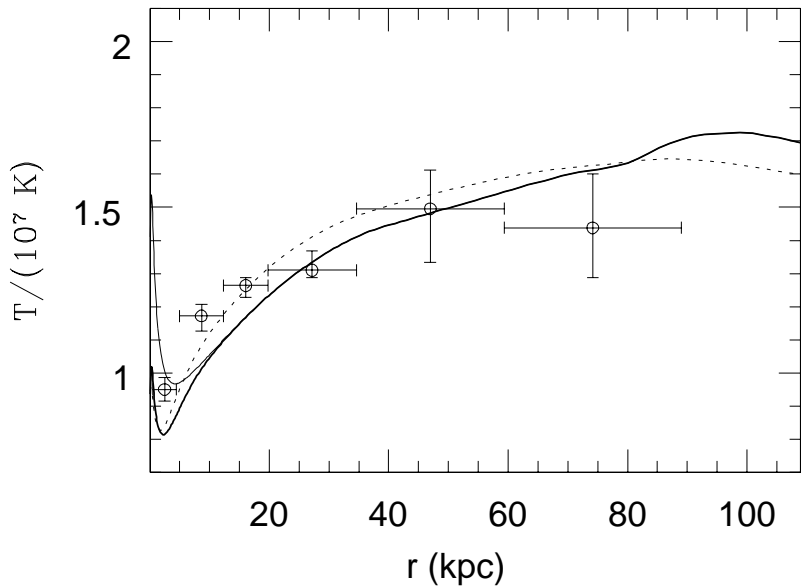
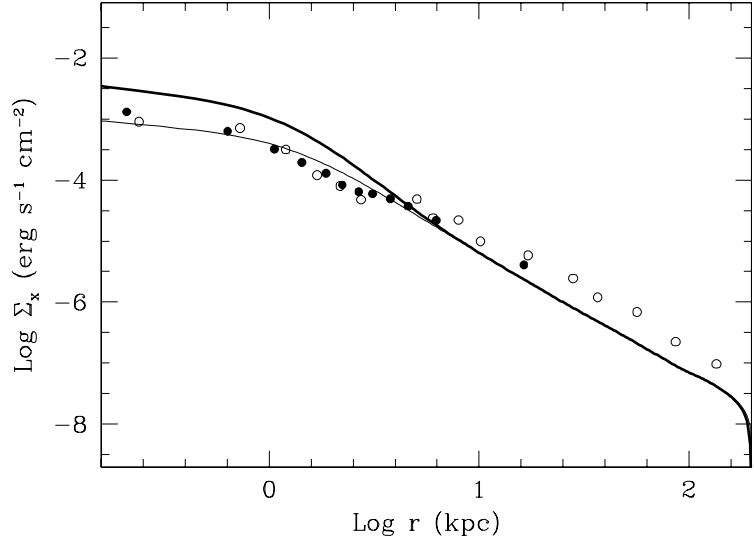
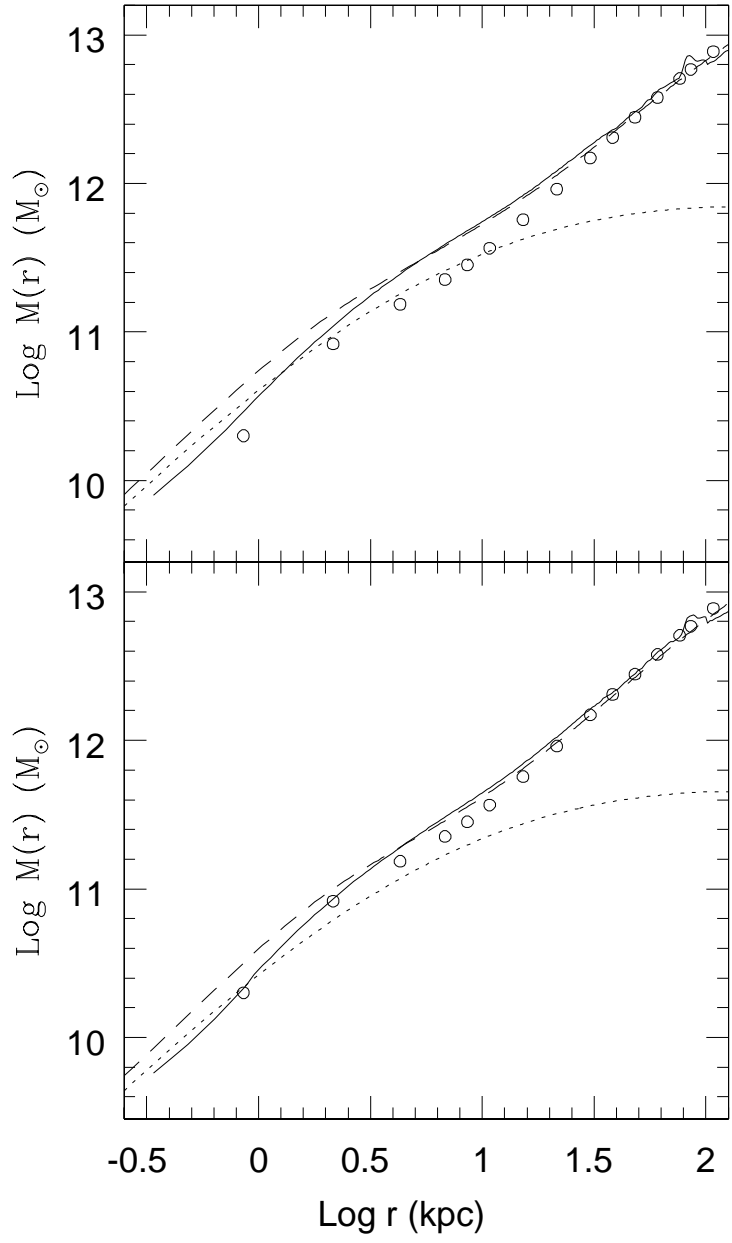
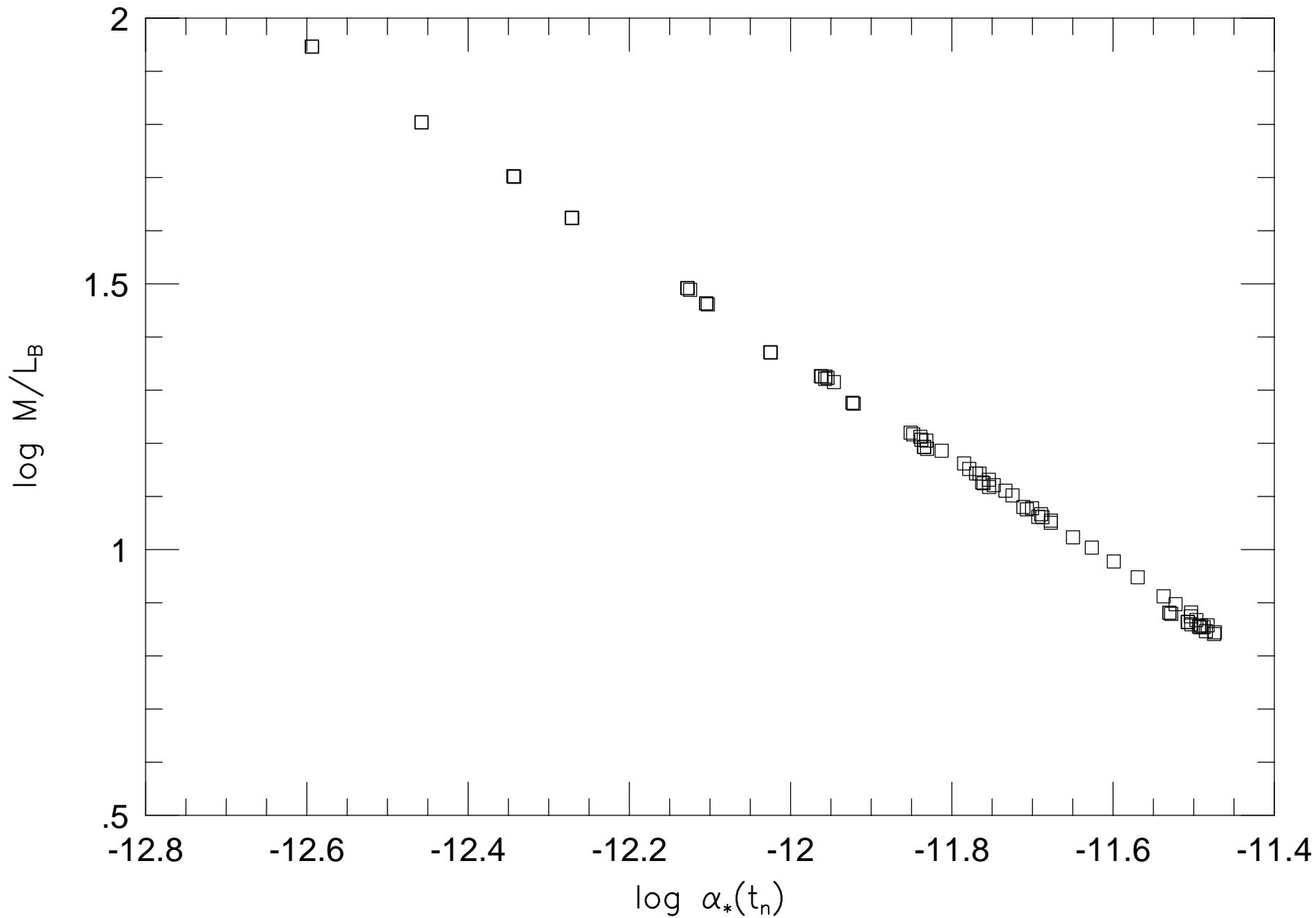
\end{document}